\documentclass[preprint, 5p, twocolumn, number, sort&compress, pdftex]{elsarticle}
\usepackage[T1]{fontenc}
\usepackage[latin1]{inputenc}
\usepackage[english]{babel}
\usepackage{amssymb, amsmath}
\usepackage{bm}
\usepackage{textcomp}
\usepackage{color}
\usepackage{lmodern}
\usepackage[bookmarksnumbered=true, bookmarksopen=false, colorlinks]{hyperref}

\newcommand{\eq}{\begin{equation}}                                                                         
\newcommand{\eqe}{\end{equation}}             
\newcommand{\pa}{\partial}

\journal{Chaos, Solitons and Fractals}
\begin{document}
\title{Self-similar analysis of a viscous heated Oberbeck-Boussinesq flow system} 

\begin{frontmatter}

\author[wigner,eli]{I.F.~Barna\corref{barnai}}

\author[sapientia]{L. M\'aty\'as}

\author[wigner,pte]{M.A. Pocsai}
\address[wigner]{Wigner Research Centre for Physics of the Hungarian Academy of Sciences, Konkoly--Thege Mikl\'os \'ut 29-33, H-1121 Budapest, Hungary}
\address[eli]{ELI-HU Nonprofit Ltd., Dugonics T\'er 13, H-6720 Szeged, Hungary}
\address[sapientia]{Department of Bioengineering, Faculty of Economics, Socio-Human Sciences and Engineering, Sapientia Hungarian University of Transylvania, Libert\u{a}tii sq. 1, 530104 Miercurea Ciuc, Romania}
\address[pte]{University of P\'ecs, Institute of Physics, Ifj\'us\'ag \'utja 6, H-7624 P\'ecs, Hungary}	

\cortext[barnai]{Corresponding author}
\ead{barna.imre@wigner.mta.hu}

\begin{abstract} 
  The simplest model to couple the heat conduction and Navier-Stokes equations  together 
is the Oberbeck-Boussinesq(OB) system which were investigated by E.N. Lorenz and opened the paradigm of chaos. 
In our former studies  -  Chaos Solitons and Fractals  {\bf{78}}, 249 (2015), 
 {\bf{ibid}}, {\bf{103}}, 336 (2017)  -  we derived analytic solutions for the velocity, 
pressure and temperature fields. Additionally, we gave a possible explanation of the 
Rayleigh-B\`enard convection cells with the help of the self-similar Ansatz.
Now we generalize the OB hydrodynamical system, including 
a viscous source term in the heat conduction equation. Our results may attract 
the interest of various fields like micro or nanofluidics or climate studies. 
\end{abstract}

\end{frontmatter}

\section{Introduction}
The investigation of coupled viscous flow equations to heat conduction has a fifteen-decade long history started with 
Boussinesq \cite{bous} and Oberbeck \cite{ober} (OB) who applied it to the normal atmosphere. 
At the beginning of the sixties - with the help of the stream function - Saltzman  \cite{salz} analyzed the problem with the help of finite Fourier series.  
At the same time Lorenz \cite{lorenz} evaluated the numerical solutions with computers and plotted the first strange attractor which was the advent of chaos as a new research field. The scientific history of this outstanding discovery can be found in the book of Gleick \cite{gleick}.  
Lorenz and Salzman both transformed the original nonlinear partial differential equation (PDE) system to a coupled nonlinear ordinary differential equation (ODE) system via a truncated Fourier series. Investigations of chaotic dynamical systems still open up new questions and help to develop new methods. 
In our first study in this field \cite{barna} we analyzed the original OB PDE system with the self-similar Ansatz ending up with a non-linear ODE system, however the pressure, temperature and velocity field was evaluated in analytic forms with the help of the error functions. As main result the possible birth of the Rayleigh-B\`enard (RB) convection cells was observed. In our second study \cite{barna2} we generalized the original OB hydrodynamical system, going beyond the first order Boussinesq approximation and consider a non-linear temperature coupling. At this point more general, power law dependent fluid viscosity or heat conduction material equations were applied. The connection of the self-similar Ansatz to critical phenomena, scaling, and  renormalization was addressed also. 

Detailed physical description and exhausted technical details about the field of RB convection can be found in numerous books \cite{ben1,ben2,ben3,ben4,ben5}.
Front propagation in RB systems - which can be investigated with the help of traveling waves - was written in detais in the review study of Saarloos \cite{saarloos}.  Pattern formation in dynamical and non-equilibrium systems is another 
relevant and never-ending research field \cite{cross} where RB convection is one of the 
most investigated phenomena \cite{rita}.  The chaotic advection phenomena can 
be properly modeled and described with the RB system as well \cite{aref}.  Additional advection phenomena in chaotic systems can be studied in
\cite{toro,boc}. The system of equations studied in \cite{barna} and \cite{barna2} 
may also contain other terms, which in the first approximation are absent 
because of the (initial, boundary, etc.) conditions, or just neglected from 
the practical point of view of the problem. 
The Navier-Stokes equation may contain couple stresses discussed in \cite{HaMe18,KhMe2014}
or the case of a transverse seepage is analyzed in \cite{AkItGe16}.   
The heat conduction equation also may contain sources or sinks, 
however a natural source term is the viscous heating \cite{GrMa1984,MaTeVo2001,TeVoMa2001}. 
Certain forms of Boussinesq description are analyzed by 
\cite{IvMe2015}.

The thermal boundary layers can be considered as a reasonable physical simplification of our present model 
and was investigated by \cite{gabi1,gabi2} with self-similar and other numerical methods.  

There is a considerable analytical and numerical effort to solve Boussinesq approximations or similar forms both for waves \cite{MaFu2006,Wa2007,Wa2008,ShKi2012,RoCh2012,Wa2012,HeSeZe2014,YaMaBa2017,KaDe2018,KoDi2012}
 and for dissipative dynamics with possible density variations \cite{DaPa2009,GaWiAu2015,An2016,LaGr2018,WeHeAh2018}. 
Experiments for certain parameter values are also realized \cite{XiZh2006,AhHeFuBo2012,AhBoHe2014}.  
Connections related to radiation and environment one may find in \cite{ PaEmPr2003}.  

The self-similar Ansatz and related constructions 
have been effectively applied in a number of hydrodynamics systems 
\cite{BaMa13,BaMa14,Yu2015,ChFaYu2017,GuUl2017,ViNaSr2018,AnPo17}. 
The book of Campos \cite{imre_book} covers more methods related to Navier-Stokes equations.   
With the help of additional Fourier transformation of the analytic velocity field connections to turbulence or enstropy could be evaluated as well. 
The references which are relevant to understand chaos and Lorenz attractor are presented in our above mentioned papers which we skip now.  
In the present study the original OB system in generalized in another way, with an 
additional viscous heating term as a source in the heat conduction equation. 
Certain nonlinear dynamical systems are also able to model the viscous heating even at the level of entropy balance \cite{MaTeVo2001}.
Beyond the self-similar Ansatz one can find other methods to solve hydrodynamics equations \cite{SaAtBa2018}. 

Viscous heating plays a crucial role in the field of micro and nanofluidics \cite{hooman,tiantian,todd}.  
Exhaustive description of viscous heating from that point of view can be found in the monographs of D. Li \cite{morini} and Gad-el-Hak \cite{hak}.   
This phenomena has relevance in other disciplines like high temperature plasma physics \cite{haines} or magma flow in geology \cite{costa}.
The mathematical properties of these kind of PDE equations attracts some interest as well, J. Li \cite{li} investigated 
the global well-posedness and formulated some theorems. 

This paper contains a self-similar analysis of the modified OB systems and the results are compared to our former ones. 
To the best of our knowledge, there is no such study available in the literature. 

\section{Theory and Results}
Let's define our field of interest as the original OB \cite{ober,salz} PDS system with the additional viscous heat source term as follows
\begin{eqnarray}
\frac{\partial u}{\partial t} + u \frac{\partial u}{\partial x} + w \frac{\pa u}{\pa z} + 
\frac{\pa P}{\pa x} -  \nonumber \\ \nu \left( \frac{\pa^2 u} {\pa x^2} +  \frac{\pa^2 u} {\pa z^2} \right)  &=& 0, \nonumber \\ 
\frac{\partial w}{\partial t} + u \frac{\partial w}{\partial x} + w \frac{\pa w}{\pa z} + \frac{\pa P}{\pa z} - eG T_1 -   \nonumber \\
 \nu \left( \frac{\pa^2 w} {\pa x^2} + \frac{\pa^2 w} {\pa z^2} \right) 
&=& 0, \nonumber \\ 
\frac{\pa T_1}{\pa t} + u \frac{\pa T_1}{\pa x}  + w \frac{\pa T_1}{\pa z} 
-     \nonumber \\ \kappa \left( \frac{\pa^2 T_1 } {\pa x^2} + \frac{\pa^2 T_1} {\pa z^2} \right) -  a \left( \frac{\pa u} {\pa z} \right)^2 &=&0,   \nonumber \\ 
\frac{\pa u}{\pa x} + \frac{\pa w}{\pa z} &=& 0,                 
\label{nav}
\end{eqnarray}
where $u,w,  $ denote respectively the x and z velocity coordinates, $T_1$ is the temperature difference relative
to the average ($T_1 = T - T_{av}$) and $P$ is the scaled pressure over the density.
There are four free physical parameters $\nu,  e, G, \kappa $ the kinematic viscosity, coefficient of 
volume expansion,  acceleration of gravitation and the coefficient of thermal diffusivity, respectively. 
(To avoid any misunderstanding we use the capital letter $G$ for gravitation acceleration and  $g$  
is reserved for a self-similar solution.)\ We denote the new parameter with $a$ which is responsible for 
the proper dimension of the viscous heating source term. 

The first two equations are the Navier-Stokes equations, the  third one is the heat conduction equation 
and the last one is the continuity equation.  All of them contain two spatial dimensions. 
We apply Cartesian coordinates and Eulerian description. 
      
We neglect the stream function reformulation of the two dimensional flow and keep the original 
variables investigating the original hydrodynamical system with the Ansatz of  
\begin{eqnarray}
u(\eta) &=& t^{-\alpha} f(\eta),  \hspace*{2mm} 
w(\eta) = t^{-\delta} g(\eta), \hspace*{2mm} \nonumber \\ 
P(\eta) &=& t^{-\epsilon} h(\eta), \hspace*{2mm} 
T_1(\eta) = t^{-\omega} l(\eta),  
\label{ans}
\end{eqnarray}
"where the new variable is  $\eta = (x+z)/t^{\beta}$. 
All exponents $\alpha,\beta,\delta,\epsilon,\omega $ are real numbers.  (Solutions with integer exponents 
are  the  self-similar solutions of the first kind and sometimes can be obtained from dimensional considerations  
\cite{sedov}.) The $f,g,h,l$ objects are called the shape functions of the corresponding dynamical variables.  
These functions should have existing first and second derivatives for the spatial coordinates and first existing 
derivatives for the temporal coordinate.  Under certain assumptions, the partial differential equations describing 
the time propagation can be reduced to ordinary differential
ones which greatly simplifies the problem. This transformation
is based on the assumption that a self-similar solution 
exists, i.e., every physical parameter preserves its shape during the expansion. Self-similar solutions usually
describe the asymptotic behavior of an unbounded or a far-field
problem; the time $t$ and the space coordinate $x$ appear
only in the combination of $x/ t^{\beta}$. It means that the existence
of self-similar variables implies the lack of characteristic
lengths and times. These solutions are usually not unique and
do not take into account the initial stage of the physical
expansion process " \cite{barna2}. 
Additional explanation of the properties of the Ansatz can be found in \cite{imre_book}. 

After some algebraic manipulation of  Eq. (\ref{nav}) all the critical exponents are fixed to the following values 
\eq
\alpha = \beta = \delta = 1/2, \hspace*{1cm} \epsilon = 1, \hspace{1cm}  
\omega = 3/2, 
\eqe
which are the same as in the "original OB" system \cite{barna} where no viscous heating term was considered. 
In the generalized case of OB  \cite{barna2} the exponents remained the same except the last one where due 
to the extra parameter the  $\lambda \cdot \omega = 3/2$ constraint is fixed. 
It is worth to mention that in the present case the fixed exponents are not enough 
to obtain an unambiguous ODE system, therefore an additional $1/\sqrt{t}$ time factor is needed to multiply 
the viscous term. Therefore, the right hand side of heat conduction equation reads $ \frac{a(u_z)^2}{\sqrt{t}}$. 
This mean, that self-similar solutions are only available when 
the heating term has to have an explicit time dependence (making the entire PDE system non-autonomous) 
and decays at large times. This property comes from the internal logic of our dispersive self-similar Anzatz and happens sometimes. 
We have to denote to our former study where the Cattaneo-Vernot telegraph heat conduction equation was 
modified to an Euler-Laplace-Darboux PDE equation which has self-similar solution with compact support. \cite{barna_heat,barna_heat2}.     

The corresponding ODE now reads the following, 
\begin{equation}\label{ode_sys}
\begin{split}
-\frac{f}{2} -\frac{f'\eta}{2} + ff' + gf' + h' -2\nu f''  &= 0,\\ 
-\frac{g}{2} -\frac{g'\eta}{2} + fg' + gg' + h' - eGl -2\nu g'' &= 0, \\ 
  -\frac{3l}{2} -  \frac{l'\eta}{2} + fl' +gl' -2\kappa l'' - a f'^2 &=0,\\
f' + g' &= 0.
\end{split}
\end{equation}

With straightforward algebraic manipulations, which were mentioned in our previous studies \cite{barna,barna2} 
well defined independent ODEs can be derived 
for the temperature, pressure and velocity shape functions. 
There is a hierarchy among the equations. In the original OB  
system the temperature is decoupled from the pressure and velocity field, 
and can be evaluated at first. Now, the hierarchy is changed and the velocity field 
became prior. 
The remaining ODE for the velocity shape function reads 
\begin{equation}
\begin{split} 
8\kappa \mu f'''' - f''' \left[ (\nu- \kappa) (4c -2\eta) \right] +   
  \\
f''\left[ 6\kappa + 2c(\eta-c) - \frac{\eta^2}{2}  +16 \right]  + 
    \\  
f'\left( \frac{\eta}{2}-c \right) + \frac{3}{2} \left(f-c \right) + eGaf'^2 = 0. 
\label{veloc}
\end{split}
\end{equation} 
Note, that if $\nu = \kappa$ which means that the kinematic viscosity and the of thermal diffusivity are equal 
(which means a very peculiar system of flow) the ODE becomes incomplete.  
As further simplification, the integration constant which comes from the continuity equation $c$ and can be set to zero. 
 
Now we get an incomplete non-linear fourth order ODE which is highly unusual. The first time derivative of 
the velocity is the corresponding acceleration which has physical interpretation but higher order time derivatives 
are meaningless in mechanical systems. (Derivation with respect to $\eta$ could be considered as time-scaled 
coordinate or space-scaled inverse time.) This fourth order ODE is originated in the couplings mechanisms of  
(\ref{ode_sys}).  

It is trivial, that without further constraints or conditions the solutions of a fourth order ODE has a very rich
 mathematical structure. The original OB system describes a fluid flow in a bounded channel, therefore 
a mixed initial and boundary problem has to be addressed. 
This condition makes the problem very similar to the Prantl boundary layer problem, which has enormous literature. 
Without completeness we mention the basic literature only \cite{schlicht}. 
The investigation of a non-Newtonian 2D laminar boundary-layer with power-law viscosity with the self-similar Ansatz leads to a non-linear 
fifth order ODE \cite{robi}. 
 
So we are interested in solutions of Eq.  (5) where the velocities and the velocity gradients are fixed at the two boundary points.  
This means that the next choice is straight forward e.g. $ f(0)  = b_1, f(\eta_1)  = d_1,  f'(0)  = b_2,  f'(\eta_1)  = d_2$. As  a  natural  choice  we  fix  the velocities  to  a  non-zero  fix  value  at  left  and  zero  value  at  the  right  boundary.   It  is  also clear that a high-order non-line ODE (which is even non-autonomous, now depends even on $\eta^2$) cannot be analyzed with a full mathematical rigor, therefore we just perform a "use your common sense" analysis and try to explore parameter sets where the evaluated solution behaves physically  reasonable. For  additionally  allowed  simplification  we  fix  the value of $e,G$ to unity and investigate the role of the viscosity $\nu$, heat conduction $\kappa$ and the strength of the viscous heating $a$ only. Figure 1 shows the shape function of the x component of the velocity between the above mentioned two boundaries for various parameter sets.  We performed numerous calculations where all three parameters $(\nu,\kappa,a)$ lie in the closed numerical range of $[0.1,10]$.  Our important experience show, that the larger the viscosity constant of the viscous heating $a$ the smaller the velocity in the chosen domain which meets our physical expectation.  For fixed viscous heating component $a$ the larger the values of $ \nu, \kappa$ the larger the velocity function in the investigated domain.The most interesting feature is the role of $c$ which is the free integration constant from the continuity equation.  Usually this value is set to zero, however for non numerical zero value the velocity shape function becomes to oscillate. The higher the $c$ value the larger the amplitudes of the oscillations.  Our explanation is the following:  higher $c$ value means higher mass flow rate, which means denser fluid, and denser fluids might have larger variations in the density (even for incompressible fluids) which is a kind of external noise.  So we think, that the numerical value of the free integration constant can be interpreted as the level of noise.  Only this parameter causes oscillations in the velocity field, otherwise the finite values of $\kappa, \nu, a $ smooth out the velocity field. Smooth velocity fields prevent the formation of  Rayleigh-B\`enard convection cells.  The main message from this study at this point is that the viscous heating term (with the finite value of $a$) prevents any kind of instability in this model.  Fig.  1 presents four different velocity fields for different parameter sets. Large 
$c$ values causes spurious oscillations.  We investigate the role of the ratios of $\kappa$ and $\nu$  if $\kappa = \nu $ the third derivative of the velocity  vanishes which simplifies the ODE.  The other two cases $(\kappa < \nu) $ and $(\kappa > \nu)$  make no difference in the final numerical results.  

For all cases presented in Fig 1 the function $f$ tends to a finite value, when $\eta \rightarrow 0$, i.e. $t\rightarrow \infty$ for finite space coordinates. Consequently the velocity function has the form 
\begin{equation} 
u \simeq \frac{const.}{t^{1/2}} ,
\end{equation}
for sufficiently large times.

Figure 2 presents the projection of the velocity function $u(x,t)$  with $c = 0$. The distribution function is a more or less flat surface with a singularity in the origin which can be  removed with the $\tilde{t} = t-t_0$ transformation.

The second independent ODE in the hierarchy is for the  shape function of the temperature field,  
\eq
 -2\kappa l'' +l'\left(c - \frac{\eta}{2}   \right)  -\frac{3l}{2}   - a f'^2 =0.\\
\label{temp}
\eqe
 Note, the direct dependence on the velocity shape field derivatives. 
 
Figure 3 shows the shape function of the temperature between the same boundaries as in Fig 1. with $c = 0$. 
The function has a clear flat minima in the investigated interval. 
As one can see on Fig. 3, as the variable $\eta=(x+z)/t^{1/2}$ decreases (finite $x+z$ and increasing time) the function $l(\eta)$ passes the zero value two times. This means that the temperature may go below the average for a while. After sufficient long time with fixed space coordinates, when $\eta$ tends to zero, the value of  $l(\eta)$ approaches a fixed value, consequently based on  (2).   
\begin{equation}
T_1 \simeq \frac{const.}{t^{3/2}} , 
\end{equation} 
which shows the long time decay of the temperature $T_1$.

Note, that the function is quick-decaying missing any kind of oscillations or additional structure. 
Figure 4 presents the projection of the complete velocity function to the x,t plane. 
Note, that the function is quick-decaying missing any kind of oscillations or additional structure.

The final ODE is for the shape function of the pressure field
\eq
h' = \frac{eGl}{2} + \frac{c}{4}. 
\label{press} 
\eqe 

Figure 5 shows this function in between two boundaries. 
The function gently increases with a smooth oscillation. 
The behavior of the pressure is relatively regular. For $\eta \rightarrow 0$ the function $h(\eta)$ tends to a fixed value as one can see on Fig 5. This means that for sufficiently long times 
\begin{equation} 
P \simeq \frac{const.}{t} 
\end{equation}
The decay of pressure is represented on Fig 6, and it has a certain monotony without oscillations.  

During our present analysis of viscous heating we were speculating about additional physically relevant heating mechanisms. It is worth to mention, that we tried to find self-similar analytic solutions for radiative heating where 
an  $a \cdot T(x,y,t)^4$ term is added to the heat conduction equation according to the well-known Stefan-Boltzmann law. The analysis of the exponents clearly showed, that an additional $t^{5/12}$ time-dependent factor is required to fulfill all necessary conditions to obtain an ODE system. Therefore think that such a term would be non-physical therefore we skip further investigation. 
To analyze the original Oberbeck-Boussinesq \cite{barna} the modified 
OB system \cite{barna2} or even the present system with the traveling-wave
Ansatz could be an additional interesting project.   

We find possible that a rotation around the y axis perpendicular to the x-z plane could be an reasonable  generalization as well. However, at first the effect of the rotation in the viscous fluid equations (without heat conduction) should be investigated and understood. 
Similar studies are already under the way. 
 
\begin{figure} 
\scalebox{0.45}{
\rotatebox{0}{\includegraphics{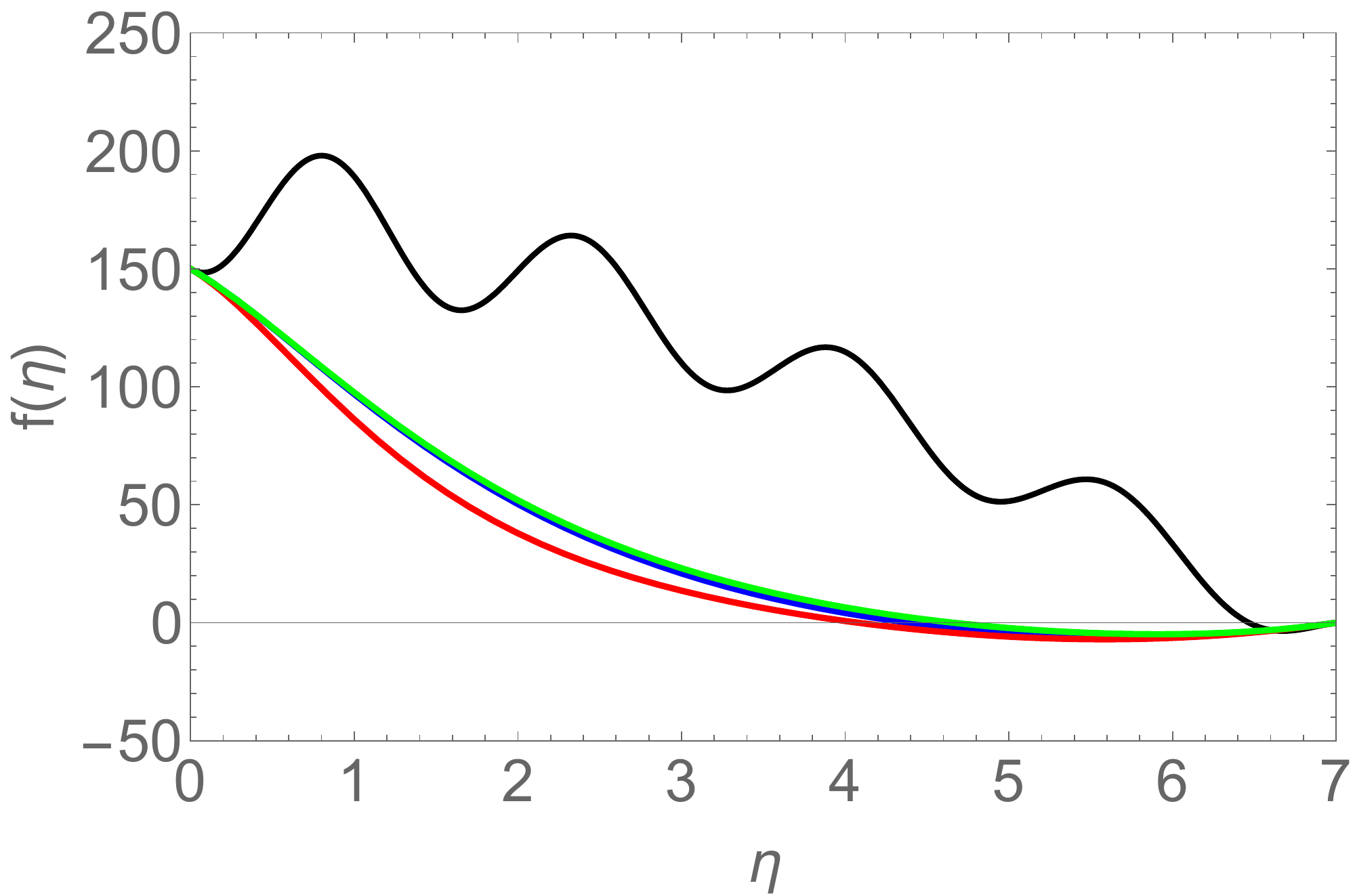}}}
\caption{The shape functions of the velocity component  $u$ as the function of $\eta$ for various physical parameter sets. 
The integration parameters are the same for all four functions $f(0) = 150, f(7) = 0, f'(0) = - 40, f'(7) = 10$ which model a flow in a finite duct. 
The velocity is maximal at one boundary and zero at the other. The acceleration is negative on the maximal velocity side and positive on the other. 
The constants $ G$ and $ e$ are set to unity and $ a = 0.5$ The black curve is for a parameter set of $\kappa = 4.1, \nu = 3.4, c =33 $. Notice the heavy oscillations. The blue, red and green curves represent results for $ c =0$ with $\kappa = \nu = 3.1$ , $\kappa = 0.8  < \nu = 4.1,$ and
$\kappa = 5.2 > \nu = 1.8,$ respectively.}
\label{f_eta}       
\end{figure}
\begin{figure} 
\scalebox{0.4}{
\rotatebox{0}{\includegraphics{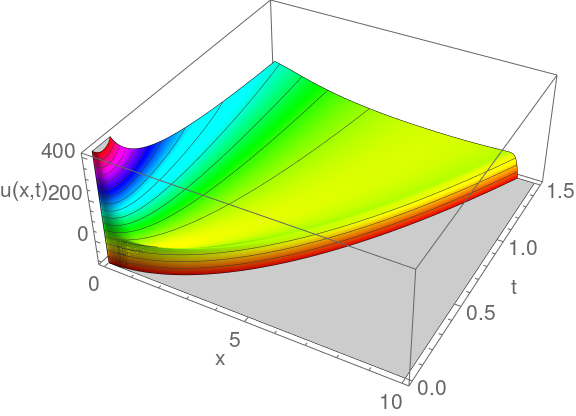}}}
\caption{The velocity distribution functions  $u(x,t) $.    The parameters are $c = 0, a = 0.5, \kappa = 0.8 $ and $\nu = 4.1$.} 	
\label{v_3d}       
\end{figure}

\begin{figure} 
\scalebox{0.45}{
\rotatebox{0}{\includegraphics{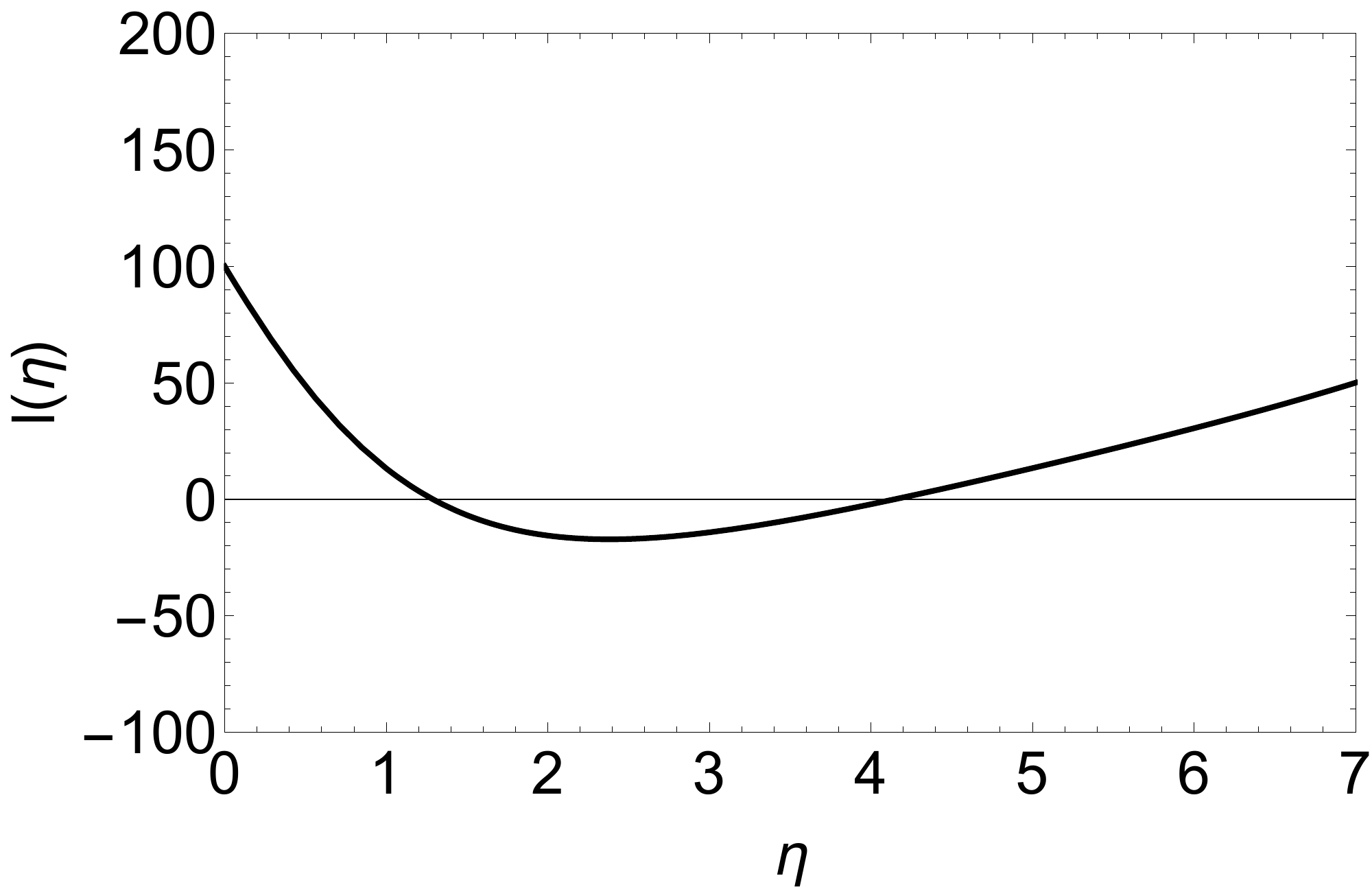}}}
\caption{The shape function of the temperature Eq. (\ref{temp}) as the function of $\eta$. The parameters are the same as at Fig. 2. }	
\label{l_eta}       
\end{figure}
\begin{figure} 
\scalebox{0.4}{
\rotatebox{0}{\includegraphics{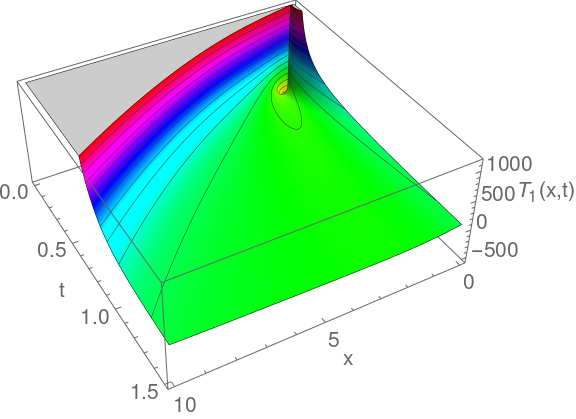}}}
\caption{The projected temperature distribution function $T_1(x,t)$. The parameters are the same as at Fig. 2. }	
\label{Txt}       
\end{figure}
\begin{figure} 
\scalebox{0.45}{
\rotatebox{0}{\includegraphics{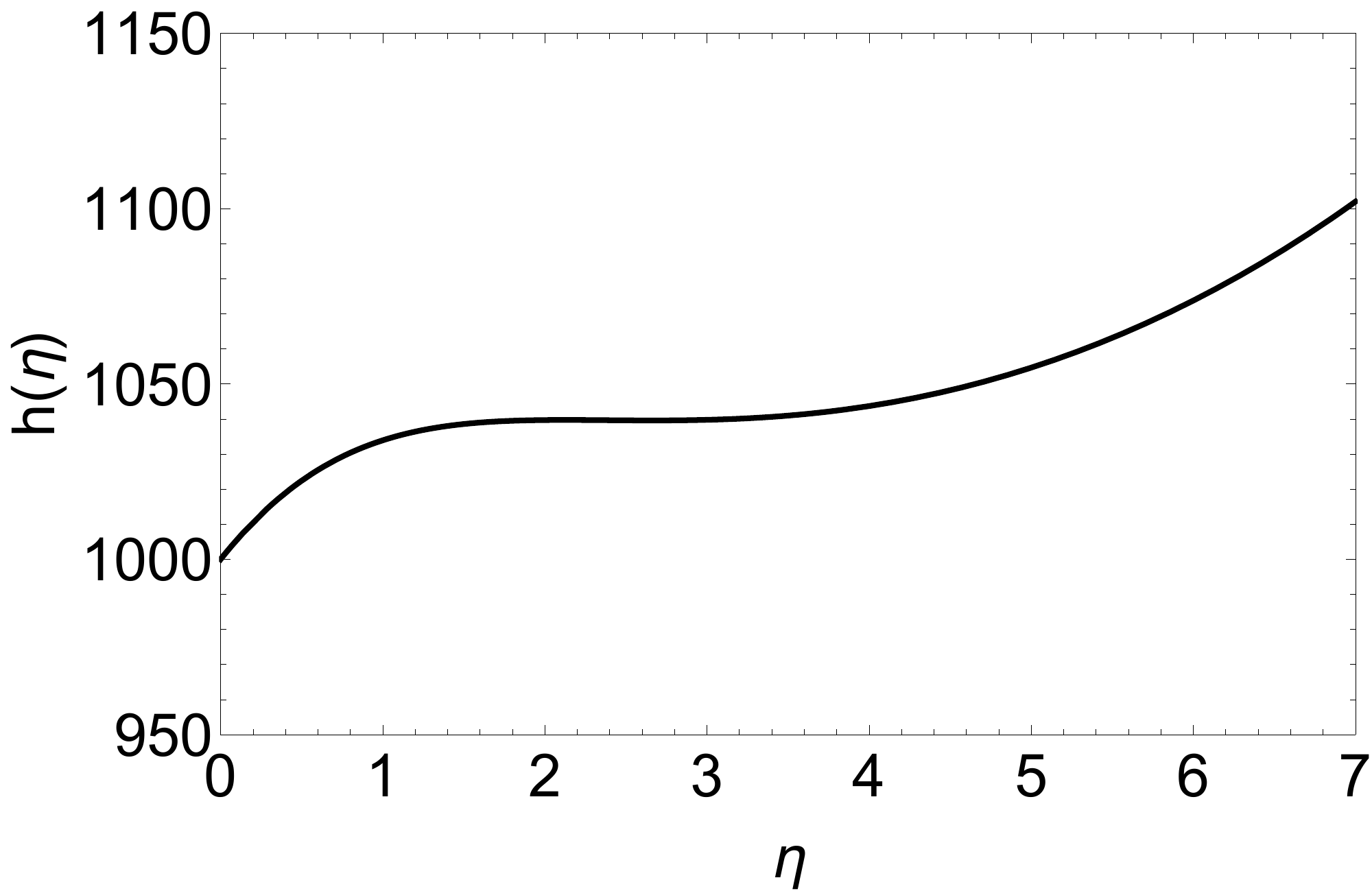}}}
\caption{The shape function of pressure field  Eq. (\ref{press}) as the function of $\eta$. The parameters are same as at Fig. 2.}	
\label{h_eta}       
\end{figure}
\begin{figure} 
\scalebox{0.4}{
\rotatebox{0}{\includegraphics{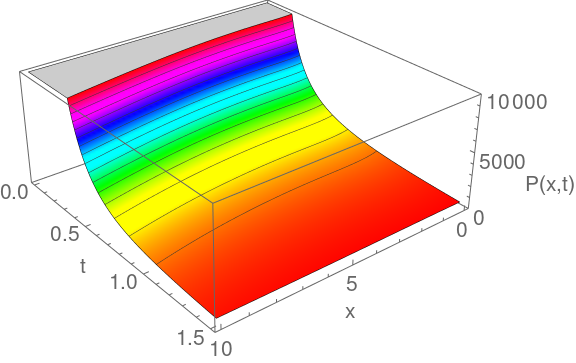}}}
\caption{The projected pressure distribution function $P(x,t)$. The parameters are same as at Fig. 2.} 	
\label{Pxt}       
\end{figure}
\section{Summary and Outlook}
We gave a physically reasonable generalization of the classical OB equation which was the first study in the 
paradigm of chaos. As a new feature we added an additional source term to the heat conduction equation, which 
is proportional to the square of the velocity gradient and called viscous heating.  Instead of the usual Galerkin 
method which applies truncated Fourier series we took the two-dimensional generalization of the self-similar 
Ansatz and found a coupled non-linear ODE system which can be solved with quadrature. 
 
To our best knowledge certain parts of the climate models are based on the OB
equations therefore our results might give an interesting contribution to such studies.

\section{Acknowledgment}

This work was supported by Project no. 129257  implemented with the
support provided from the National Research, Development and
Innovation Fund of Hungary, financed under the OTKA 2018 funding
scheme. The ELI-ALPS project (GINOP-2.3.6-15-2015-00001) is supported by 
the European Union and co-financed by the European
Regional Development Fund.

 

\end{document}